\documentclass[10pt,conference]{IEEEtran}

\usepackage{graphicx}
\usepackage{comment}
\usepackage{mathtools}
\usepackage{blindtext}
\usepackage{xcolor}
\usepackage{float}
\usepackage{url}
\usepackage{xspace}
\usepackage[ruled]{algorithm2e}

\newcommand{\PYP}{\texttt{pycparser}}
\newcommand{\C}{{Csmith}}
\newcommand{\K}{\texttt{K-Means}}
\newcommand{\G}{{GCC}}

\newcommand{\Comment}[1]{}
\newcommand{\Space}[1]{}
\newcommand{\test}{test input}
\newcommand{\Test}{Test input}

\newcommand{\Part}[1]{\textit{#1}}
\usepackage{fancybox}
\newcounter{observation}
\newcommand{\observation}[1]{\refstepcounter{observation}
        \begin{center}
        \Ovalbox{
        \begin{minipage}{0.93\columnwidth}
                \textbf{Observation \arabic{observation}:} #1
        \end{minipage}
        }
        \end{center}
        \vspace{-5pt}
}

\newcommand{\bug}{failure-inducing test input}
\newcommand{\KC}{\textsc{K-Config}\xspace}

\title{K-CONFIG: Using Failing Test Cases to Generate Test Cases in GCC Compilers}
\author{
    \textbf{Md Rafiqul Islam Rabin}\\
    University of Houston\\
    mdrafiqulrabin@gmail.com
    \and
    \textbf{Mohammad Amin Alipour}\\
    University of Houston\\
    amin.alipour@gmail.com
}

\begin{document}

\maketitle

\begin{abstract}

The correctness of compilers is instrumental in the safety and reliability of other software systems, as bugs in compilers can produce programs that do not reflect the intents of programmers. Compilers are complex software systems due to the complexity of optimization. \G{} is an optimizing C compiler that has been used in building operating systems and many other system software.

In this paper, we describe \KC, an approach that uses the bugs reported in the \G{} repository to generate new \test{}s. Our main insight is that the features appearing in the bug reports are likely to reappear in the future bugs, as the bugfixes can be incomplete or those features may be inherently challenging to implement hence more prone to errors. Our approach first clusters the failing \test{} extracted from the bug reports into clusters of similar \test{}s. It then uses these clusters to create configurations for \C{}, the most popular test generator for C compilers. 
In our experiments on two versions of \G{}, our approach could trigger up to 36 miscompilation failures, and 179  crashes, while \C{} with the default configuration did not trigger any failures. This work signifies the benefits of analyzing and using the reported bugs in the generation of new \test{}s. 

\end{abstract}

\section{Introduction}

\begin{figure*} [ht]
    \centering
    \includegraphics[width=0.9\linewidth]{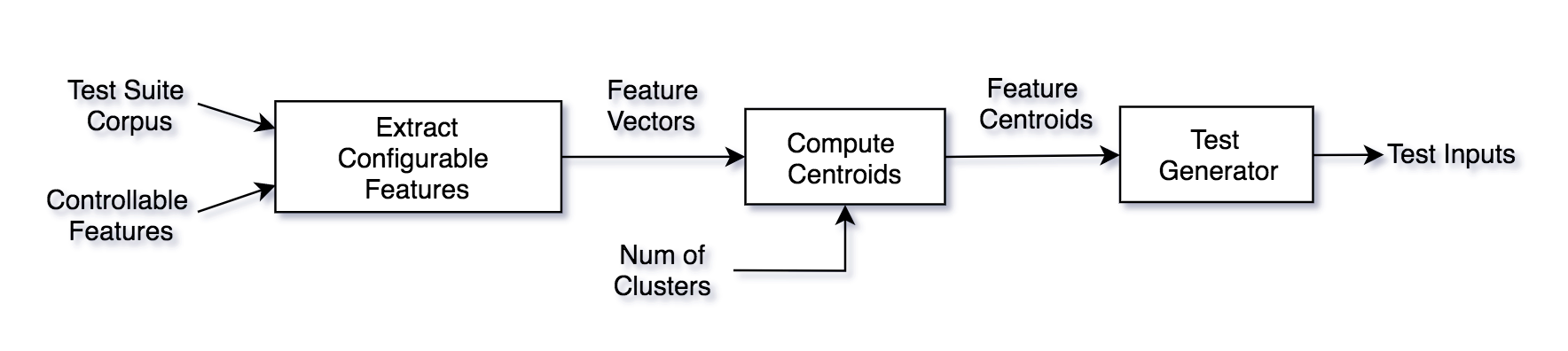}
    \caption{The overall workflow of \KC approach.}
    \label{fig:workflow}
\end{figure*}

Compilers translate programs understandable by developers to programs that machines can understand and execute.
Compilers are the key part of software development infrastructure that makes  all software systems depend on them. 
Developers \emph{rely} on compilers to build and debug their programs, libraries, and operating systems.
Optimization passes in the compilers search the input programs for the opportunities to improve various aspects of the output programs such as execution time, memory consumption, and the code size.

Optimizing compilers are complex software systems that constitute several passes from various syntax and semantic analyses to code generation. As programming languages grow and add new features, the compilers that implement these features also grow in size and complexity. Moreover,  compilers attempt to accommodate the translation of several programming languages which further complicates the compiler system. Today's \G{} compiler is over 10 million lines of code \cite{Link:GCCLOC}. Testing such massive, sophisticated systems is a non-trivial task and researchers and developers still can find many bugs in modern compilers.

Several approaches for testing compilers have been proposed; for example, \cite{Zeller:TestGenMutation:LangFuzz:Security:2012}, \cite{Regehr:TestSelection:Rank:PLDI:2013}, \cite{Zhendong:TestOracle:classfuzz:PLDI:2016}, \cite{Cummins:TestGenLearn:DeepSmith:ISSTA:2018}, \cite{Zhendong:TestGenMutation:Orion:PLDI:2014}, \cite{Zhendong:TestGenMutation:Athena:OOPSLA:2015}, \cite{Zhendong:TestGenMutation:Hermes:OOPSLA:2016}, \cite{Groce:TestGenRnd:Swarm:ISSTA:2012}, to name few. These approaches either generate \test{}s from scratch by grammar \cite{Regehr:TestGenGrammar:Csmith:PLDI:2011} and learning \cite{Cummins:TestGenLearn:DeepSmith:ISSTA:2018}, or they create new \test{} by manipulating \cite{Zeller:TestGenMutation:LangFuzz:Security:2012} or transforming the existing \test{}, e.g., \cite{Zhendong:TestGenMutation:Orion:PLDI:2014}. 

In this paper, we evaluate the use of failure-inducing \test{}s to generate new \test{}s. 
Our insight is that these \test{}s can provide hints into places in the code that are more prone to be buggy. In fact, this idea is not that novel. 
LangFuzz~\cite{Zeller:TestGenMutation:LangFuzz:Security:2012} transplants fragments of failing \test{}s to other programs to generate new \test{}. However, our work takes a significantly different approach. In this approach that we call \KC, instead of embedding fragments for existing failing test inputs into new \test{} to create a new \test{}, we analyze features of failing \test{}s to create \emph{new configurations} for a test generator. The test generator uses these configurations for creating new \test{} that exhibits similar features to the original failing \test{}.
This approach is also different from (deep) learning-based approaches such as DeepSmith~\cite{Cummins:TestGenLearn:DeepSmith:ISSTA:2018}, whereas they try to build a generative model for the test inputs in two ways. First, while learning approaches requires many test inputs with millions of tokens to train a model, this approach can work with a couple of thousands of failing test inputs. Second, learning based approaches tend to converge to a limited language model of \test{} that overly restricts the type of \test{}s that can be produced
 \cite{Godefroid:FeatureBased:LearnFuzz:ASE:2017}. \KC instead uses the configuration of test generators to guide testing which is less constrained than the generation of \test{}s in learning-based approaches. In particular, \KC only specifies the programming constructs that should be present in the generated \test{}s, and the order or number of those constructs are determined by the test generator. 
 

Figure~\ref{fig:workflow} depicts the overall workflow of the approach. It constitutes following main phases: (1) collecting failing \test{}s, (2) extracting configurable test features from failing \test{}s, (3) clustering \test{}s into a similar cluster, (4) generating configurations based on clusters, and finally (5) using configurations to generate new \test{}. 

Of course, there are limitations to the application of this approach. First, it assumes that a stable test generator exists. Second, it requires a set of failing test inputs. \G{} compiler easily satisfies both requirements, as it has been under development for decades and the bug reports are available. Moreover, it has a mature well-engineered test generator, \C{}~\cite{Regehr:TestGenGrammar:Csmith:PLDI:2011} \cite{Link:Csmith}. It allows us to evaluate the effectiveness of this approach in testing \G{} compilers.

We have implemented the proposed approach for GCC C compiler testing. We collected 3661 failing \test{}s from \G{} codebase. We parse the \test{}s and collect the features that can be used in the configuration of \C{}. We used the \K{} algorithm to cluster the \test{}s. \K{} returns centroids of the clusters of similar \test{}s. We use these centroids to synthesize configurations for \C{}.

We performed a large scale experiment to evaluate the effectiveness of configurations generated by this approach with the default configuration of \C{} on two versions of \G{}. In total, we experimented the new and default configuration for over 900 hours (almost 40 days). The result of our experiment shows that the new configurations could find up to 36 \test{} for miscompilation, and 179 \test{} for crashes per 13-hour test sessions, while \C{} with the default configuration could not find any failures at the same time.
It reinforces the previous studies that many of bugfixes are incomplete~\cite{Yin:IncompleteBugFix:ESEC:FSE:2011} and \G{} is not an exception. This also indicates that processing failing \test{} can provide insights into the regions of code that are susceptible to bugs.


\textbf{Contributions}
This paper makes the following main contributions:
\begin{itemize}
    \item We propose a novel approach for testing compilers with mature test generators.
    \item We perform a large-scale study to evaluate the efficiency of the proposed approach.
    \item We make code and \test{} available for further use.  \footnote{We add the URL to the data and code at the time of publication.}
\end{itemize}

\textbf{Paper Organization}
Section~\ref{sec:approach} demonstrates the proposed approach. Section~\ref{sec:setting} describes the experimental setup for the evaluation of the approach.
Section~\ref{sec:analysis} provides an analysis of the results and answer to the research questions. 
Section~\ref{sec:related} surveys the related works, Section~\ref{sec:threat} discusses some of the threats to validity. Finally, Section~\ref{sec:conclution} concludes the paper.

\section{Proposed Approach}
\label{sec:approach}

Main programming languages such as C and JavaScript have test generators that produce \test{}s for those languages.
The \test{}s can be used to test compilers and interpreters of the languages. 
For example, \C{}  has been able to find hundreds of bugs in mainstream C compilers.
Another good example is jsFunFuzz~\cite{Link:jsFunFuzz} for JavaScript that has found thousands of bugs in JavaScript interpreters.
Newer programming languages are also developing such tools for testing compilers, for example, Go-Fuzz~\cite{Link:goFuzz} for Go.

Configurable test generators, such as \C{}, allow developers to specify some of the characteristics of the \test{} to be generated. 
This way developers can control the test generation and direct the test generation process.
It is fair to say that the current test generation techniques under-utilize the configuration of the test generators.
We only could find two studies that use the configurations: swarm testing~\cite{Groce:TestGenRnd:Swarm:ISSTA:2012} and focused random testing~\cite{Alipour:FocusedTesting:ISSTA:2016}. 
Swarm testing randomly enables or disables options in the test generators.
Focused random testing attempts to establish a causation relation between configurable options and test coverage in order to find configurations that can target parts of the code.

\newcommand{\TS}{$TS$\xspace}
\newcommand{\TG}{\textsc{TestGen}\xspace}

In this section, we describe the proposed approach in detail. Our approach is based on analysis of previous failing \test{}s to generate the configuration for the test generators that we call it \KC.
\KC takes \TS, a set of existing \test{}s that exhibit some interesting property $P$, and a configurable test generator \TG for a compiler.
The goal of \KC is to analyze \TS to extract $k$ configurations for \TG that are likely to generate test cases that exhibit $P$.

\begin{figure} [!ht]
    \centering
    \includegraphics[width=\columnwidth]{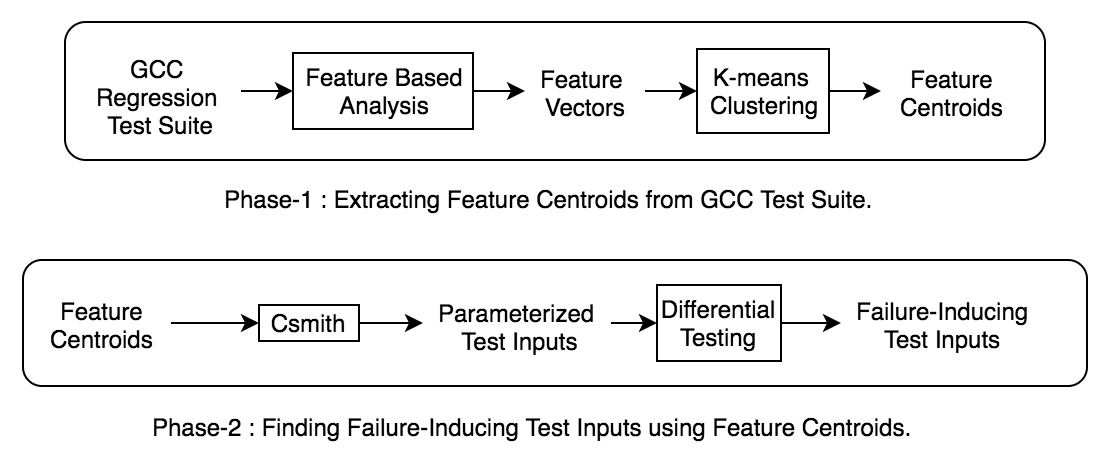}
    \caption{Overview of \KC approach for testing \G{}.}
    \label{fig:process}
\end{figure}

Figure ~\ref{fig:process} depicts an overview of the workflow for realizing \KC for testing the \G{} compiler.
It can broadly be divided into two phases: (1) extracting feature centroids from test suite of failing \test{}s, and (2) using centroids to generate test cases. We describe these phases in the following subsections.


\subsection{Phase 1: Extracting Feature Centroids from Test Suite}

\Part{Configurable Test Generator}
\C{}~\cite{Regehr:TestGenGrammar:Csmith:PLDI:2011} is a configurable random test generator for C compilers.
The common practice for testing compilers is differential testing. That is a \test{} is compiled and executed by two or more versions of compilers, or two or more optimization levels, and the results are compared.
The metamorphic test oracle specifies the result of the output of the compiled programs by all compilers and optimization levels must be the same. 
An obnoxious feature of testing compilers, especially C compilers is that the language allows undefined behavior. 
Undefined behaviors are those the standard of the language does specify standard behavior of the program for certain conditions. 
For example, the C standard does not specify default values for the uninitialized variables in the programs; it, therefore, is up to the developers of the compilers to decide on the actual behavior. \C{} does the best effort to avoid the undefined behaviors in C.

\Part{Controllable Features}
\C{} allows developers to choose the C programming constructs that they want in the \test{}s generated by \C{}. The order and number of the constructs however are chosen randomly and developers cannot control them---mainly because \C{} is a random generator that uses grammar to generate test cases. 
We use \C{} 2.3.0 in our experiments that provides 28 configuration options. 
These options are offered in the form of ``\texttt{feature}'' for including the \texttt{feature} in the \test{} and ``\texttt{no-feature}'' to exclude the feature in the generation of the \test{}s.
For example, \C{} includes a \texttt{volatile} variable  in \test{} by  using ``\texttt{--volatiles}'' or excludes it by using ``\textit{--no-volatiles}''.
The 28 features list: \textit{``argc, arrays, bitfields, comma-operators, compound-assignment, consts, divs, pre-incr-operator, pre-decr-operator, post-incr-operator, post-decr-operator, unary-plus-operator, jumps, longlong, int8, uint8, float, inline-function, muls, packed-struct, pointers, structs, unions, volatiles, volatile-pointers, const-pointers, global-variables, and builtins"}

\Part{Test Suite Corpus}
At first, to extract the properties of the failing \test{}, we need a corpus of failing test suite. 
We extracted $7131$ \test{} from bug reports in \G{}; these \test{}s caused some older versions of \G{} to fail.
We use these \test{}s as the basis of our analysis.

\Part{Extracting Test Features}
We use \PYP{} v2.19 \cite{Link:PyCParser}, a parser for the C programming language written in Python to extract C programming constructs used in the test suite corpus. 
We use the abstract syntax tree (AST) to find out features present in the \test{}s in the test suite. 
An unanticipated finding was that the \PYP{} failed on C programs having comments. It is therefore likely that \PYP{} failed on \G{} regression test C programs as those C programs contain comments. To resolve this issue and make \PYP{} working on those C programs, we removed the comments from C programs. Finally, \PYP{} was able to parse 3,661 of 7,131 test C programs. We investigated the rest of the \test{} and found that they are indeed not parsable, but they caused the earlier versions of \G{} to crash. 
We used 3,661 parsable \test{} and their corresponding AST in our experiments.

\begin{figure*} [ht]
    \centering
    \includegraphics[width=0.9\linewidth]{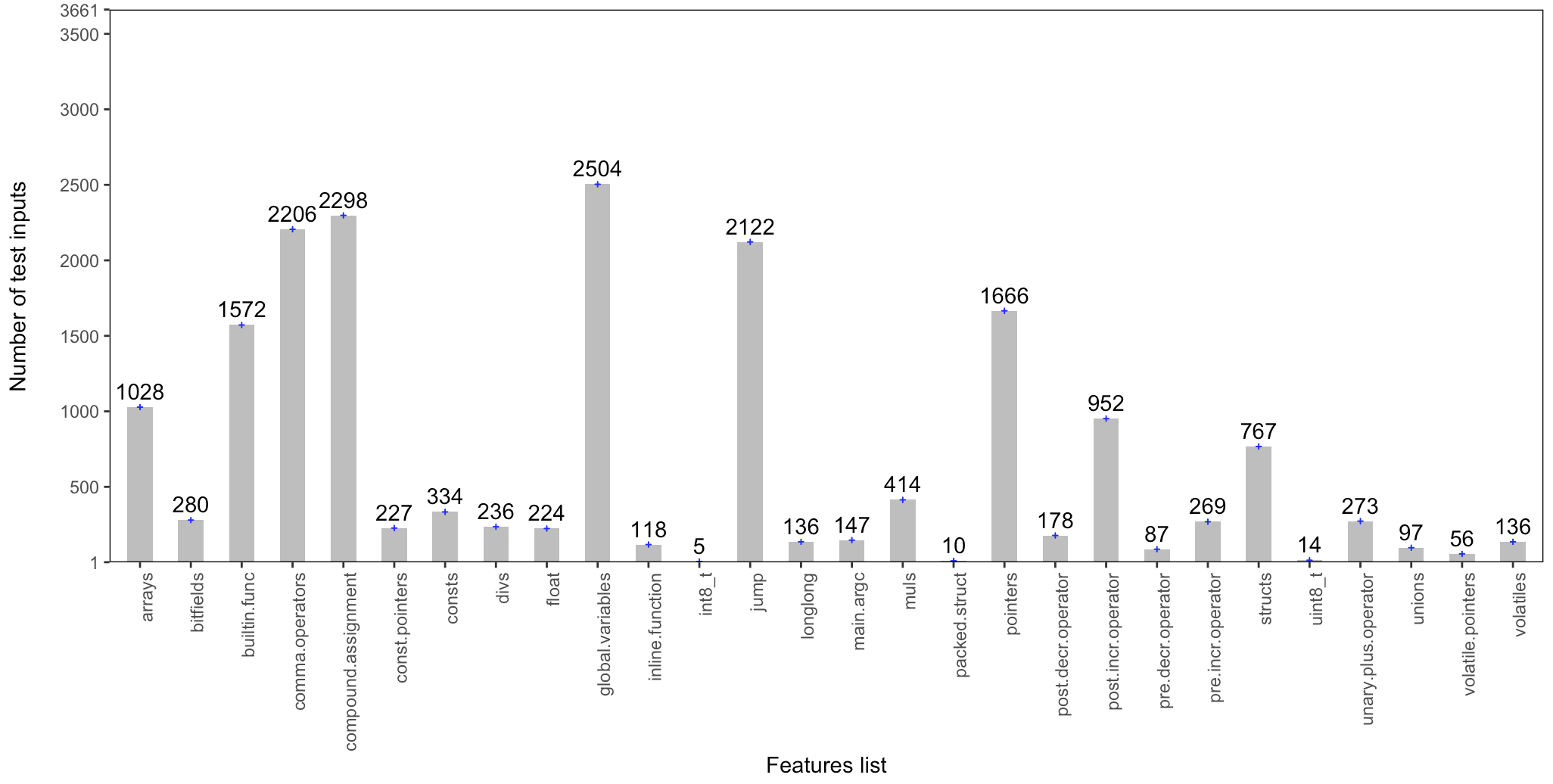}
    \caption{Number of \test{} for each feature.}
    \label{fig:feature_count_testinputs}
\end{figure*}

\Part{Extracting Feature Vectors}
In this step, we focus on counting the number of  occurrences of each of the 28 features in our test suite. 
To do this, we use a combination pattern matching in regular expressions in the text of \test{}s and simple visiting of the abstract syntax tree. We extract all occurrences of features in \test{}s. 
We next count the number of occurrences for each feature in each C program or corresponding AST file.
Figure ~\ref{fig:feature_count_testinputs} shows the number of \test{} that contains each feature. We observe that the distribution of features in failing \test{} is not uniform. Features like global variables and compound assignments 
have occurred more frequently than features like int\_8. 

\Part{Compute Centroids}
Given a set of feature vectors that represents the presence or absence of each feature in the \test{}s, 
we use \K{} clustering on the feature vectors.
The \K{} clustering is an unsupervised machine learning algorithm that performs clustering on unlabeled vector data. 
Given a set of vector data, this algorithm observes the underlying patterns and cluster similar data together. 
The number of clusters we want to see has to be predefined. 
Each cluster results in a \emph{centroid} that has a minimum distance to the data points of the cluster. 
Suppose, $V$ is the vector data of $n$ observations and $k$ is the number of disjoint clusters $C$. The \K{} algorithm groups the $n$ observations into $k$ clusters and each cluster has a centroid $c$, the mean of the samples $V_c$ in the cluster. The centroid $c$ is set based on the minimum distance $d_m$ of the inertia criterion. For \K{}, the distance metric is the sum of squared distances within-cluster which is defined as: 
\begin{align*}
  d_m = \sum_{i=0}^{n}\min_{c_i \in C, x_j \in V_{c_i}}(|c_i - x_j|^2)
\end{align*}

\K{} computes $k$ centroids for a given $k$.
At the end of \K{} clustering, we have $k$ clusters and $k$ centroids where are located at the center of each cluster.
Since the feature vectors contain only 0 and 1 values, the values in the centroids would be a real value from 0 to 1 (inclusive).

\subsection{Phase 2: Generating \test{} using \KC} 

\Part{Generating configurations in \KC}
Our realization of \KC uses \K{} that results in a vector of real values from 0 to 1.
A closer value to 1 in a centroid, it means that the corresponding feature was more prevalent in the \test{}s in that cluster. Therefore, we use those values as the probability of including a feature in a \test{}. Algorithm~\ref{alg:generator} describes the algorithm for generating new \test{}s. 
Given a testing time budget $timeBudget$, a set of centroids $CS$, the algorithm calls $ConfigGen$ in round-robin fashion until the test time budget expires.  
Procedure $ConfigGen$ takes a centroid $C\in CS$ and generates a new configuration. 
In generating a new configuration, $ConfigGen$ chooses to include feature $f_i$ with a random probability $c_i$ where $f_i$ is represented by the element $c_i$ in the centroid vector.

\begin{algorithm} [ht]
\caption{\KC}
\label{alg:generator}
 $timeBudget$  $\leftarrow$ Testing time budget\;
 $CS$ $\leftarrow$ Set of centroids\;
 $TS \leftarrow \emptyset$\;
 \DontPrintSemicolon\;
\While{$spentTime \leq timeBudget$}{
 \ForAll{centroid $C$ $\in$ $CS$}{
     $config \leftarrow ConfigGen(C)$\;
     $testInput \leftarrow Csmith(config)$\;
     \If{doesFail(testInput,GCC)}{
       $TS \leftarrow TS \cup testInput$\;
     }
 }
}
\DontPrintSemicolon\;
\SetKwProg{Fn}{Function}{:}{\KwRet}
\Fn{ConfigGen($C$)}{
    $features \leftarrow \emptyset$\;
    \ForAll{value $v$ $\in$ $C$}{
      $randNum \leftarrow [0:1]$\;
    \eIf{randNum $\leq$ v}{
        $features.put(1)$\;
    }{
        $features.put(0)$\;
    }
 }
}
\end{algorithm}

\Part{Differential testing to evaluate test input}
We use a metamorphic relation between optimization levels of compilers.
In particular,  we compile a \test{} with two optimization levels: \texttt{O0} and \texttt{O3} and we compare the result of the execution of the programs generated by those optimization levels. 

\section{Experimental Setup}
\label{sec:setting}
This section discusses the experimental parameters used to evaluate the \KC approach.

\begin{figure} 
    \centering
    \includegraphics[width=\columnwidth]{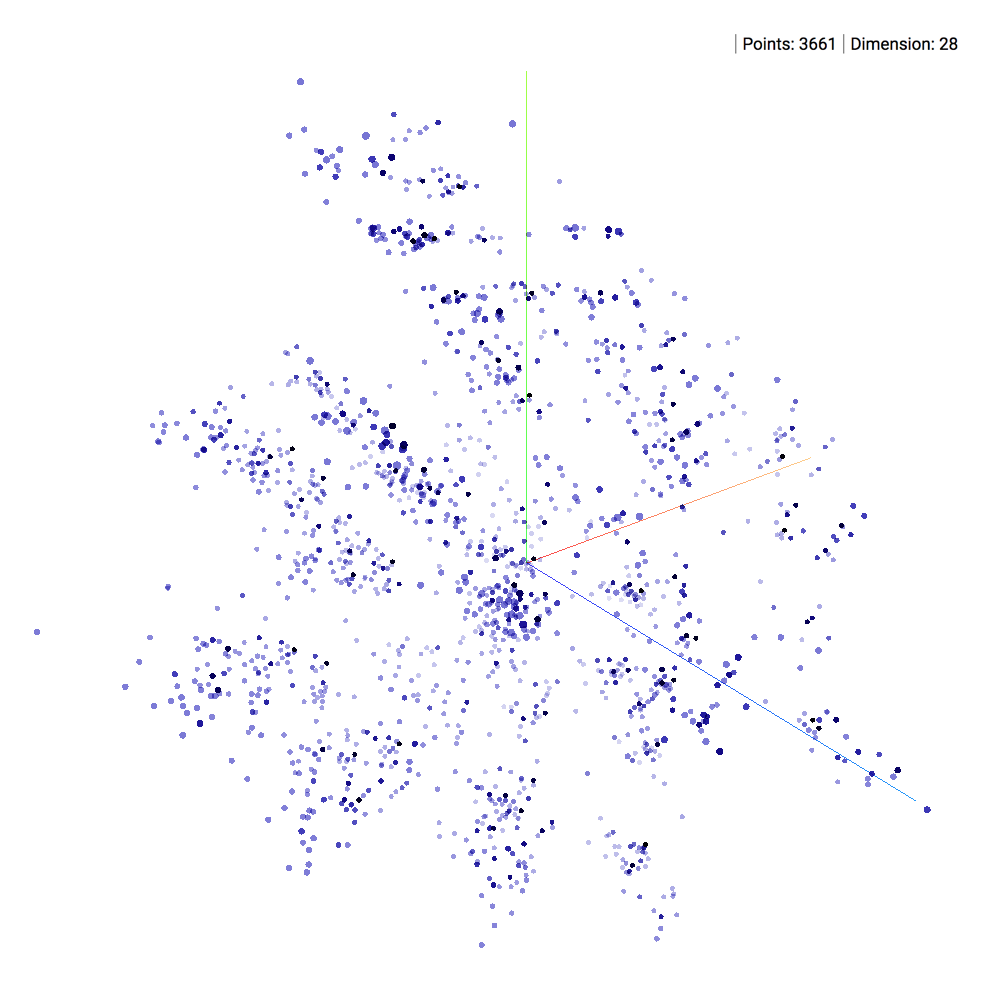}
    \caption{Visualization of feature vectors of failing \test{}s in \G{}.}
    \label{fig:feature_vector_projection}
\end{figure}

\Part{Number of Clusters}
Choosing the number of cluster $k$ is key and hard. The best $k$ ensures the similarity within the clusters and dissimilarity between the clusters. But there are no well-defined methods to choose such a value of $k$. We visualize our feature vectors in the projector \cite{Link:ProjectorTensorflow} to see the underlying clustering patterns. Figure ~\ref{fig:feature_vector_projection} shows the projection of our vector data. Here, the number of points is $3661$ and the dimension of each point is $28$. That means each point is the representation of a test input having a vector of 28 features. We can see different cluster patterns in various aspects, that's why we come up with a decision to choose different $k$ values, where $k = {1,2,4,8,16}$.

\subsection{Initial Setting for Clusters}
We consider the \K{} algorithm implementation of python \texttt{scikit-learn} \cite{Link:ScikitLearn} machine learning library in this paper. We use the \texttt{KMeans} API where we need to fit the vector data (the observations to the cluster) and have to pass the \texttt{n\_clusters} parameters (the number of centroids to generate after forming the number of clusters). We also use the default \texttt{\K{}++} initialization method which selects initial cluster centers for \K{} clustering in a smart way to speed up the convergence. The inertia criterion for distance metric is used is the sum of squared distances between centroid and data points. After each iteration, the \K{} algorithm minimizes the within-cluster sum of squared distances. We run our algorithms with default \texttt{n\_int} and \texttt{max\_iter} option. As a result, the \K{} algorithm runs 10 times with different centroid seeds and continue for 300 iterations for each run. The centroids are found at the last iteration of \K{} that dumps the best output as the final result.

\Part{Test Subject Compilers}
We use two mature versions of \G{} to evaluate the effectiveness of this approach: \G{} 4.8.2 and \G{} 5.4.0. \G{} 4.8.2 was released in October of 2013 and \G{} 5.4.0 was released three years after, in September 2016. Both releases are mature and have been widely used in building various software systems.

\Part{System Hardware}
Our evaluation has been conducted on a high performance computing cluster. The HPC Server is the shared campus resource pool hosting a total of 5704 CPU cores in 169 computes. The CPU type is Intel Xeon E5-2680v4 with 128GB shared main memory.

\Part{Test Generation Tool}
\C{} \cite{Regehr:TestGenGrammar:Csmith:PLDI:2011} is an open source automatic test generation tool. Given a set of C language features as options (by default enable), \C{} can generate random C programs. We use \C{} 2.3.0 \cite{Link:Csmith} in our approach.

\Part{Initial Test Suite}
We use the \G{} regression bug test suite that has more than 3000 parsable test C programs. This test suite contains failure-inducing \test{}s. We are interested in mine the patterns of those failure-inducing \test{} to guide the \C{} \test{} generation.

\Part{Test budget for testing campaign}
We run each configuration of \C{} for 13 hours to create \test{} and execute the \test{}. We also experiment with two compilation time setup. First, we use a 10 seconds timeout to compile a \test{}. Then, we use a 30 seconds timeout to compile a \test{}. To check the robustness of randomness, we run each experiment three times.

\subsection{Research Questions}

In this study, we seek to answer the following research questions. 

\begin{itemize}
    \item Research Question 1: Can the \KC find more \bug{}s compared to the state-of-the-art approach?
    \item Research Question 2: What impact of choosing the different $k$ have on \KC?
    \item Research Question 3: What are the common features in the failure-inducing \test{}s for \G{}?
\end{itemize}

For \textbf{RQ1}, we seek to compare the \C{} with our setting to the \C{} with the default setting in terms of the number of failure-inducing \test{} for the crash, timeout, and miscompilation. \textbf{RQ2} evaluates the impact of choosing different $k$ on the effectiveness of \KC. Finally, for \textbf{RQ3}, we aim to find the features that are culprits for the failures.


The remaining of this section discusses our evaluation of the proposed \KC approach. We have been experimenting on \G{}.

\subsection{Ground Truth}
To find the \bug{}s, we choose the result of without optimization as ground truth. For example, for a specific compiler version, we first compiled a \test{} with the lowest optimization (-O0). Then, we compiled the same \test{} with the highest optimization (-O3) on the same compiler version.
Test oracles state that the behavior should be the same in both trials. Any mismatch between the behaviors represents a failure.
There are other failures that are described in Table ~\ref{table:problematic_combinations}. 

\begin{table}
    \begin{center}
        \caption{Possible failures in the experiment}
        \label{table:problematic_combinations}
        \resizebox{\columnwidth}{!}{%
        \begin{tabular}{|c|c|c|}
            \hline
            \textbf{No Optimization (-O0) } & \textbf{High Optimization (-O3)} & \textbf{Failure?} \\
            \hline \hline 
                Compiler crashes & Compiler crashes & False \\ \hline
                Compiler crashes & Compiler doesn't crash & True \\ \hline
                Compiler doesn't crash & Compiler crashes & True \\ \hline
                \multicolumn{2}{|c|} {Output for -O0 and -O3 are identical} & False \\ \hline
                \multicolumn{2}{|c|} {Output for -O0 and -O3 are different} & True \\ 
            \hline
        \end{tabular}%
        }
    \end{center}
\end{table}

\subsection{Failure Types}
We have classified the failures into three classes: (1) miscompilation, (2) crash failure, and (3) timeout. 
Table ~\ref{table:failure_types} summarizes these failures.
The miscompilation failures happen wherein a compiler produces programs that output wrong outputs for different optimization. In a crash failure, the compiler terminates the compilation abruptly with a crash report on screen. 
The timeout failure happens when the compilation time exceeds the predefined threshold for compilation---we need to set timeout due to avoid potential infinite loop errors in the compiler under test.

\begin{table}
    \begin{center}
        \caption{Failure types}
        \label{table:failure_types}
        \resizebox{\columnwidth}{!}{%
        \begin{tabular}{|c|l|}
            \hline
            \textbf{Failure Types} & \textbf{Definition} \\
            \hline \hline 
                Miscompilation & Compiler produces different output for no (-O0) and high (-O3) optimization. \\ \hline
                Crash & Compiler crashes when compiling program. \\ \hline
                Timeout & Compiler takes longer time than the specified time to compile program. \\
            \hline
        \end{tabular}%
        }
    \end{center}
\end{table}

\subsection{Experiment parameters}
Table ~\ref{table:experiment_types} contains information about the experiments.
We have conducted six experiments to evaluate the effectiveness of the \KC approach. 
In E1, we used \C{} with the default configuration. 
In E2 through E6 experiments, instead of using the default setting of \C{}, 
we used the different featured centroids as parameters for \C{}. 
We chose different $k$ values, where $k = {1,2,4,8,16}$, respectively. 
$k=1$ is essentially setting the probability of including a feature proportional to the number of times that is it has seen in \test{}s in the original test suite. 
In each experiment, for 13 hours, we generated, compiled \test{}, and executed the output programs.
We also ran each experiment three times to avoid potential effects of randomness in the experiments.  

\begin{table}
    \begin{center}
        \caption{Experiment with GCC for different feature selection}
        \label{table:experiment_types}
        \resizebox{\columnwidth}{!}{%
        \begin{tabular}{|c|c|l|c|}
            \hline
            \textbf{Experiment ID} & \textbf{Feature Selection} & \textbf{Testing Window} \\
            \hline \hline 
                E1 & Csmith default configuration & 13 hours \\ \hline
                E2 & Select k=1 centroid as features & 13 hours \\ \hline
                E3 & Select k=2 centroids as features & 13 hours \\ \hline
                E4 & Select k=4 centroids as features & 13 hours \\ \hline
                E5 & Select k=8 centroids as features & 13 hours \\ \hline
                E6 & Select k=16 centroids as features & 13 hours \\ \hline
        \end{tabular}%
        }
    \end{center}
\end{table}

\section{Analysis of results}
\label{sec:analysis}

This section presents the results of experiments to evaluate the effectiveness of the \KC approach. 
Tables \ref{table:gcc482_t10} and \ref{table:gcc482_t30} show the experiment results with \G{} 4.8.2 for the compilation timeout of 10 seconds and 30 seconds respectively. 
Table \ref{table:gcc540_t10} and \ref{table:gcc540_t30} show the experiment results with \G{} 5.4.0 for the compilation time of 10 seconds and 30 seconds respectively.

\begin{table*}
    \begin{center}
        \caption{GCC 4.8.2 (timeout=10s)}
        \label{table:gcc482_t10}
        \resizebox{\textwidth}{!}{%
        \begin{tabular}{|c|c|c|c|c|c|c|c|c|c|c|}
            \hline
            \textbf{Experiment ID} & \textbf{\Test{}} & \textbf{Crash(0)} & \textbf{Crash(3)} & \textbf{Crash(both)} & \textbf{Total Crash} & \textbf{Timeout(0)} & \textbf{Timeout(3)} & \textbf{Timeout(both)} & \textbf{Total Timeout} & \textbf{Miscompilation} \\
            \hline \hline
            E1 (r1)&9225&0&0&0&0&8&0&1187&1195&0 \\
            E1 (r2)&10037&0&0&0&0&10&0&1226&1236&0 \\
            E1 (r3)&9075&0&0&0&0&13&0&1164&1177&0 \\
            \hline
            E2 (r1)&10351&54&83&7&144&21&12&1545&1578&26 \\
            E2 (r2)&11524&63&102&13&178&29&18&1609&1656&25 \\
            E2 (r3)&10122&66&95&6&167&28&18&1509&1555&24 \\
            \hline
            E3 (r1)&10577&52&71&7&130&20&17&1542&1579&31 \\
            E3 (r2)&10952&49&76&5&130&17&8&1669&1694&22 \\
            E3 (r3)&10262&56&70&5&131&24&12&1512&1548&20 \\
            \hline
            E4 (r1)&11325&48&62&3&113&19&11&1531&1561&24 \\
            E4 (r2)&11835&62&79&3&144&22&16&1634&1672&28 \\
            E4 (r3)&10897&49&80&7&136&24&11&1499&1534&26 \\
            \hline
            E5 (r1)&10411&49&73&10&132&24&13&1596&1633&23 \\
            E5 (r2)&10777&48&83&4&135&24&10&1714&1748&26 \\
            E5 (r3)&9717&38&69&4&111&16&12&1586&1614&21 \\
            \hline
            E6 (r1)&11024&60&115&4&179&32&12&1535&1579&30 \\
            E6 (r2)&11449&61&103&6&170&28&21&1649&1698&29 \\
            E6 (r3)&10754&50&94&10&154&28&12&1496&1536&36 \\
            \hline
        \end{tabular}%
        }
    \end{center}
\end{table*}

\begin{table*}
    \begin{center}
        \caption{GCC 5.4.0 (timeout=10s)}
        \label{table:gcc540_t10}
        \resizebox{\textwidth}{!}{%
        \begin{tabular}{|c|c|c|c|c|c|c|c|c|c|c|}
            \hline
            \textbf{Experiment ID} & \textbf{\Test{}} & \textbf{Crash(0)} & \textbf{Crash(3)} & \textbf{Crash(both)} & \textbf{Total Crash} & \textbf{Timeout(0)} & \textbf{Timeout(3)} & \textbf{Timeout(both)} & \textbf{Total Timeout} & \textbf{Miscompilation} \\
            \hline \hline
            E1 (r1)&8285&0&0&0&0&8&0&1059&1067&0  \\
            E1 (r2)&8722&0&0&0&0&9&0&1137&1146&0  \\
            E1 (r3)&8225&0&0&0&0&11&0&1010&1021&0  \\
            \hline
            E2 (r1)&9901&0&0&0&0&5&1&1471&1477&41  \\
            E2 (r2)&10637&0&0&0&0&5&1&1560&1566&51  \\
            E2 (r3)&9565&0&0&0&0&3&0&1438&1441&42  \\
            \hline
            E3 (r1)&10044&0&0&0&0&5&1&1481&1487&38  \\
            E3 (r2)&10407&0&0&0&0&8&1&1591&1600&39  \\
            E3 (r3)&9768&0&0&0&0&4&3&1443&1450&32  \\
            \hline
            E4 (r1)&10224&0&0&0&0&7&0&1449&1456&42  \\
            E4 (r2)&10895&0&0&0&0&4&2&1525&1531&28  \\
            E4 (r3)&9633&0&0&0&0&3&1&1430&1434&30  \\
            \hline
            E5 (r1)&9541&0&0&0&0&6&0&1514&1520&45  \\
            E5 (r2)&10020&0&0&0&0&5&0&1624&1629&35  \\
            E5 (r3)&9413&0&0&0&0&5&1&1478&1484&30  \\
            \hline
            E6 (r1)&10306&0&0&0&0&4&2&1436&1442&53  \\
            E6 (r2)&10588&0&0&0&0&5&3&1544&1552&47  \\
            E6 (r3)&9915&0&0&0&0&6&1&1399&1406&37  \\
            \hline
        \end{tabular}%
        }
    \end{center}
\end{table*}

\begin{table*}
  \begin{center}
    \caption{GCC 4.8.2 (timeout=30s)}
    \label{table:gcc482_t30}
    \resizebox{\textwidth}{!}{%
    \begin{tabular}{|c|c|c|c|c|c|c|c|c|c|c|}
     \hline
      \textbf{Experiment ID} & \textbf{\Test{}} & \textbf{Crash(0)} & \textbf{Crash(3)} & \textbf{Crash(both)} & \textbf{Total Crash} & \textbf{Timeout(0)} & \textbf{Timeout(3)} & \textbf{Timeout(both)} & \textbf{Total Timeout} & \textbf{Miscompilation} \\    
        \hline \hline
        E1 (r1)&4400&0&0&0&0&3&0&552&555&0 \\
        E1 (r2)&4832&0&0&0&0&4&0&609&613&0 \\ 
        E1 (r3)&4316&0&0&0&0&6&0&540&546&0 \\ 
        \hline
        E2 (r1)&4516&19&38&3&60&9&4&620&633&7 \\
        E2 (r2)&4637&25&49&2&76&11&2&685&698&12 \\ 
        E2 (r3)&4361&32&45&1&78&17&6&606&629&12 \\ 
        \hline
        E3 (r1)&4252&29&26&4&59&4&4&633&641&12 \\
        E3 (r2)&4550&23&37&1&61&6&4&688&698&12 \\ 
        E3 (r3)&4221&19&26&3&48&8&3&618&629&12 \\ 
        \hline
        E4 (r1)&4786&19&30&3&52&7&5&616&628&13 \\
        E4 (r2)&4944&26&26&1&53&12&6&677&695&18 \\ 
        E4 (r3)&4406&19&34&4&57&10&4&607&621&13 \\ 
        \hline
        E5 (r1)&4456&19&37&2&58&15&4&627&646&10 \\
        E5 (r2)&4370&16&27&2&45&5&5&695&705&11 \\ 
        E5 (r3)&4232&25&33&2&60&5&4&623&632&15 \\ 
        \hline
        E6 (r1)&4372&26&39&3&68&16&3&618&637&12 \\
        E6 (r2)&5140&19&49&0&68&15&4&674&693&21 \\ 
        E6 (r3)&4219&21&34&3&58&13&7&611&631&15 \\ 
        \hline
    \end{tabular}%
    }
  \end{center}
\end{table*}

\begin{table*}
  \begin{center}
    \caption{GCC 5.4.0 (timeout=30s)}
    \label{table:gcc540_t30}
    \resizebox{\textwidth}{!}{%
    \begin{tabular}{|c|c|c|c|c|c|c|c|c|c|c|}
     \hline
      \textbf{Experiment ID} & \textbf{\Test{}} & \textbf{Crash(0)} & \textbf{Crash(3)} & \textbf{Crash(both)} & \textbf{Total Crash} & \textbf{Timeout(0)} & \textbf{Timeout(3)} & \textbf{Timeout(both)} & \textbf{Total Timeout} & \textbf{Miscompilation} \\    
        \hline \hline
        E1 (r1)&4110&0&0&0&0&5&0&515&520&0 \\
        E1 (r2)&4587&0&0&0&0&2&0&573&575&0 \\ 
        E1 (r3)&4270&0&0&0&0&5&0&505&510&0  \\
        \hline
        E2 (r1)&4189&0&0&0&0&1&0&614&615&15 \\
        E2 (r2)&4648&0&0&0&0&4&0&667&671&15 \\ 
        E2 (r3)&4062&0&0&0&0&1&0&609&610&5  \\
        \hline
        E3 (r1)&4132&0&0&0&0&0&0&615&615&23 \\
        E3 (r2)&4385&0&0&0&0&2&0&676&678&14 \\ 
        E3 (r3)&4273&0&0&0&0&0&0&597&597&10  \\
        \hline
        E4 (r1)&4333&0&0&0&0&2&0&605&607&16 \\
        E4 (r2)&4603&0&0&0&0&1&0&669&670&16 \\ 
        E4 (r3)&4420&0&0&0&0&0&0&594&594&26  \\
        \hline
        E5 (r1)&3949&0&0&0&0&1&0&616&617&17 \\
        E5 (r2)&4090&0&0&0&0&1&0&683&684&11 \\ 
        E5 (r3)&4114&0&0&0&0&4&0&602&606&19  \\
        \hline
        E6 (r1)&4198&0&0&0&0&0&0&615&615&25 \\
        E6 (r2)&4707&0&0&0&0&2&0&663&665&19 \\ 
        E6 (r3)&4003&0&0&0&0&1&0&599&600&21  \\
        \hline
    \end{tabular}%
    }
  \end{center}
\end{table*}

Here, the first column shows the type of experiments. Column ``Test input'' shows the total number of \test{}s generated and executed in 13 hours period for a specific experiment. 
``Crash(0)'' shows the total number of crash failures that have been found in an experiment while compiling a \test{} with the lowest optimization level (i.e., -O0), ``Crash(3)'' demonstrates the number of crashes encountered with the highest level of optimization (i.e., -O3). 
``Crash(both)'' column contains the number of \test{} that causes a crash in both lowest (-O0) and highest (-O3) level of optimization. 
``Total Crash'' is the sum of ``Crash(0)'', ``Crash(3)'', and ``Crash(both)'' in an experiment. 
Similarly, ``Timeout(0)'', ``Timeout(3)'', and ``Timeout(both)'' represent the number of timeouts encountered with lowest (-O0) optimization, highest (-O3) optimization, and both the lowest (-O0) \& highest (-O3) optimization, respectively. 
``Total Timeout'' is the sum of ``Timeout(0)'', ``Timeout(3)'', and ``Timeout(both)''. 
Note that we run each experiment three times; (r1), (r2), and (r3) present the result of individual experiments.

\subsection*{\textbf{RQ1}: Comparison of the configuration of \KC with the default configuration of the test generator}

In \G{} 4.8.2, the configuration of the \KC approach could find up to 179 \test{} for crashes and 36 \test{} for miscompilation failures. 
But \C{} with the default configuration could not find any failure-inducing \test{}. On the other hand, in \G{} 5.4.0, \KC suggests configurations that could find up to 53 \test{} with miscompilations and no \test{} for crashes. Again \C{} with the default setting could not find any failure-inducing \test{}. 

\observation{\C{} with \KC configurations could find more failure-inducing \test{}s than \C{} with the default configuration.}

\subsection*{\textbf{RQ2}: What impact of different $k$ effectiveness of \KC}

We experiment with a different selection of $k$ in our approach. Table \ref{table:gcc482_t10}-\ref{table:gcc540_t30} summarizes the results of those selections. When \C{} with default selection could not find any failure-inducing \test{}, \C{} with $k=1$ setting even could find many failure-inducing \test{}. $k=1$ is similar to set the probability of configurations based on the ratio of the number of test inputs per feature. Whatever value $k$ takes we could see the robustness of finding failure-inducing \test{} in each experiment. Therefore, choosing the $k$ value has no major impact on \KC.

\observation{Different values for $k$  did not impact the effectiveness of \KC.}

\begin{figure*} [ht]
    \centering
    \includegraphics[width=0.9\linewidth]{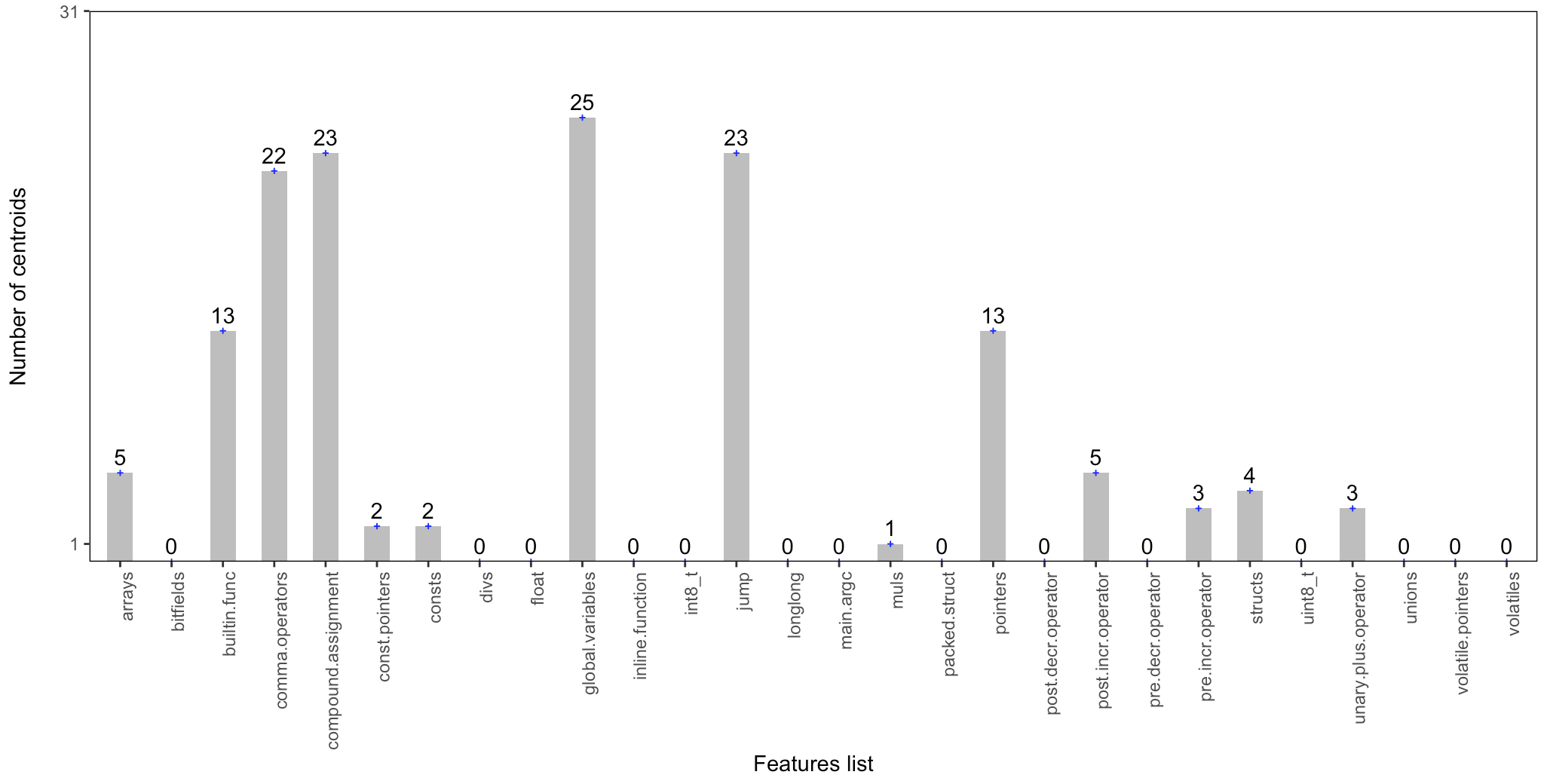}
    \caption{Frequency of features in centroids.}
    \label{fig:feature_count_centroids}
\end{figure*}

\subsection*{\textbf{RQ3}: Common features in the failure-inducing \test{}s for \G{}}

To get rid of bias for centroids in the experiment, we run round-robin selection of centroids in \test{} generation. Figure ~\ref{fig:feature_count_testinputs} shows the quantitative count of each feature in the \test{}. We could observe that there is no missing feature in the test suite, the feature exists in at least 5 to 2504 \test{}, which supports the goodness of our test suite. Similarly, Figure ~\ref{fig:feature_count_centroids} shows the quantitative count of each feature in centroids. We could observe that \textit{``global-variables, compound-assignment, jumps, comma-operators, pointers, builtins"} features appear exceedingly in the failure-inducing \test{}s, and \textit{``arrays, post-incr-operator, structs, pre-incr-operator, unary-plus-operator, consts, const-pointers, muls"} features appear occasionally in the failure-inducing \test{}s. On the other hand, \textit{``argc, bitfields, divs, pre-decr-operator, post-decr-operator, longlong, int8, uint8, float, inline-function, packed-struct, unions, volatiles, volatile-pointers"} features appear rarely in the failure-inducing \test{}s. 

\observation{Table \ref{table:common_features} shows the number of occurrences of the common features in the failure-inducing \test{}s.}

\begin{table}
    \begin{center}
        \caption{Ranking of features based on their frequency in failure inducing \test{}s}
        \label{table:common_features}
        \resizebox{\columnwidth}{!}{%
        \begin{tabular}{|c|l|}
            \hline
            \textbf{Frequency of features} & \textbf{Feature list} \\
            \hline \hline 
                Very frequent & \textit{``global-variables, compound-assignment, jumps, comma-operators,} \\ & \textit{pointers, builtins"}
                \\ \hline
                Occasionally & \textit{``arrays, post-incr-operator, structs, pre-incr-operator, unary-plus-operator,} \\ & \textit{consts, const-pointers, muls"}
                \\ \hline
                Rarely & \textit{``argc, bitfields, divs, pre-decr-operator, post-decr-operator, longlong, int8,} \\  & \textit{uint8, float, inline-function, packed-struct, unions, volatiles, volatile-pointers"}
                \\ \hline
        \end{tabular}%
        }
    \end{center}
\end{table}

\section{Related Work}
\label{sec:related}

Finding bugs in compilers is an active area of research. Several approaches have been proposed  to generate test inputs, reduce test size, rank and prioritize test inputs, select/omit/mutate features, diverse test inputs, accelerate testing, categorize similar bugs, etc. The goals of those approaches are to reveal bugs in compiler and facilitate debugging. This section highlights the literature of compiler testing related to our proposed approach. The summary of related works is shown in Table ~\ref{table:related_works}.

\begin{table*} [ht]
    \begin{center}
    \caption{Summary of related works}
    \label{table:related_works}
    \resizebox{\textwidth}{!}{%
    \begin{tabular}{c | c | c | c}
    \hline \hline
    Author and Reference & Publication Year & Approach(s) & Test Compiler/Engine \\
    \hline
    Leon et. al. \cite{Leon:TestSelection:coverage:ISSRE:2003}
    & ISSRE 2003 & Filtering and prioritizing & GCC, Jikes and javac \\
    Podgurski et. al. \cite{Podgurski:BugTriage:profile-visualization:ICSE:2003}
    & ICSE 2003 & Classification and visualization & GCC and Jacks \\
    Leon et. al. \cite{Leon:TestSelection:CIF:ICSE:2005}
    & ICSE 2005 & Filtering with cif & javac, Xerces and JTidy \\
    Holler et. al. \cite{Zeller:TestGenMutation:LangFuzz:Security:2012}
    & USENIX 2012 & Fuzzing code fragments & Mozilla TraceMonkey, Google V8 and PHP \\
    Groce et. al. \cite{Groce:TestGenRnd:Swarm:ISSTA:2012}
    & ISSTA 2012 & Feature omission & LLVM/Clang and GCC \\
    Chen et. al. \cite{Regehr:TestSelection:Rank:PLDI:2013}
    & PLDI 2013 & Ranking test inputs & SpiderMonkey and GCC \\
    Zhang et. al. \cite{Zhendong:TestSelection:SPE:PLDI:2017}
    & PLDI 2017 & Skeletal program enumeration & GCC/Clang, CompCert, Dotty and Scala \\
    Chen et. al. \cite{Chen:TestSelection:LET:ICSE:2017}
    & ICSE 2017 & Learning and scheduling & GCC and LLVM \\
    Chen et. al. \cite{Chen:TestSelection:COP:IEEE-TSE:2018}
    & IEEE TSE 2018 & Coverage prediction and clustering & GCC and LLVM \\
  \hline \hline
  \end{tabular}%
  }
  \end{center}
\end{table*}

\subsection{Clustering-based approaches}
Cluster filtering approaches have been used in the compiler testing for classifying test inputs, triggered bugs, and failure reports. This classification can help researchers to pick test programs from the different cluster in order to increase the diversity of program as well the testing acceleration as testing from the same cluster is likely to observe the same facts. 
Several approaches for cluster filtering have been proposed; for example, \cite{Leon:TestSelection:coverage:ISSRE:2003}, \cite{Podgurski:BugTriage:profile-visualization:ICSE:2003}, \cite{Leon:TestSelection:CIF:ICSE:2005}, \cite{Chen:TestSelection:COP:IEEE-TSE:2018}.
These approaches either cluster based on the profile execution (\cite{Leon:TestSelection:coverage:ISSRE:2003}, \cite{Podgurski:BugTriage:profile-visualization:ICSE:2003}) or filter based on the information flow (\cite{Leon:TestSelection:CIF:ICSE:2005}, \cite{Chen:TestSelection:COP:IEEE-TSE:2018}).
Filtering and Prioritizing \cite{Leon:TestSelection:coverage:ISSRE:2003} partitions a set of test cases into separate groups according to the profile similarity on execution space. The authors then use the one-per-cluster sampling to select one test program randomly from each cluster. If any bug found, the authors use the failure-pursuit sampling to select $k$ nearest neighbors of the failure-inducing program. This process is continued until no more bugs are found. Another work in this area is classification and multivariate visualization \cite{Podgurski:BugTriage:profile-visualization:ICSE:2003} where the failure-inducing inputs have been grouped together based on the profile execution space. The classification approach has four phases: (1) capturing the execution profile in the first phase, (2) extracting profile features, (3) grouping similar failures using cluster analysis and multivariate visualization, and (4) explore the results in order to confirm or refine. Filtering with complex information flows \cite{Leon:TestSelection:CIF:ICSE:2005} is another profile-based test case filtering approach where both coverage-based and profile-distribution-based filtering approaches are considered. Another example is COP \cite{Chen:TestSelection:COP:IEEE-TSE:2018} where authors prioritize test inputs by clustering them according to the predicted coverage information. The evaluation result of all those approaches demonstrates that the cluster filter approach is effective and finds many defects along with maximizing the coverage.

\subsection{Feature-based Testing}
Feature selection leads a program generator to generate diverse test programs that explore the various area of space. These diverse test inputs can increase code coverage and find hidden bugs. 
Several feature-based analysis have been proposed; for example, \cite{Groce:TestGenRnd:Swarm:ISSTA:2012}, \cite{Zhendong:TestSelection:SPE:PLDI:2017}, \cite{Zeller:TestGenMutation:LangFuzz:Security:2012}, \cite{Godefroid:FeatureBased:LearnFuzz:ASE:2017}.
 Swarm testing \cite{Groce:TestGenRnd:Swarm:ISSTA:2012} randomly chooses a subset of features available to generate new test cases. The generated test cases are very diverse and the evaluation result shows that this approach outperforms \C{}'s default configuration in both code coverage and crash bug finding. Another notable work is SPE \cite{Zhendong:TestSelection:SPE:PLDI:2017} where authors enumerate a set of programs with different variable usage patterns. The generated diverse test cases exploit different optimization and the evaluation result shows that the skeletal program enumeration has confirmed bugs in all tested compilers. Two more related studies in this area are LangFuzz \cite{Zeller:TestGenMutation:LangFuzz:Security:2012} and Learn\&Fuzz \cite{Godefroid:FeatureBased:LearnFuzz:ASE:2017}. LangFuzz approach extracts code fragments from a given code sample that triggered past bugs and then apply random mutation within a pool of fragments to generate test inputs. On the other hand, the Learn\&Fuzz approach uses the generative learned char-RNN model to generate new test objects for the experiment. 

\subsection{Accelerate Testing}
Another important fact behind the diverse test input is to accelerate the testing. Running a larger set of test inputs will take a long period of time to find compiler bugs, and repeatedly testing similar test inputs will result in the same compiler bugs. 
Several approaches for testing acceleration have been performed; for example, \cite{Regehr:TestSelection:Rank:PLDI:2013}, \cite{Chen:TestSelection:LET:ICSE:2017}, \cite{Chen:TestSelection:4steps:ICSE:2018}, \cite{Chen:TestSelection:COP:IEEE-TSE:2018}.
One novel approach is taming \cite{Regehr:TestSelection:Rank:PLDI:2013} where authors order the test inputs in such a way that diverse test inputs are highly ranked. They first define distance metrics between test cases and then rank test cases in the furthest point first order. The evaluation result shows that the ranking approach speeds up the bug finding in both the number of test inputs and testing time. Another interesting approach is LET \cite{Chen:TestSelection:LET:ICSE:2017} where authors use a learning model to schedule the test inputs. This learning-to-test approach has two steps: learning and scheduling. In learning steps, LET extracts a set of features from the past bug triggering test cases and then trains a capability model to predict the bug triggering probability of the test programs, and trains another time model to predict the execution time of the test programs. In scheduling steps, LET ranks the target test programs based on the probability of bug triggering in unit time. The evaluation result shows that the scheduled test inputs significantly accelerate compiler testing. 
Another example in this area is COP \cite{Chen:TestSelection:COP:IEEE-TSE:2018} where authors predict the coverage information of compilers for test inputs and prioritize test inputs by clustering them according to the predicted coverage information. The result shows that COP significantly speeds up the test acceleration and outperforms the state-of-the-art acceleration approaches.

\section{Threats to Validity}
\label{sec:threat}

There are various factors that may impact the validity of our results.
First, our initial test suite of failing \test{}s is not representative of all bugs in the GCC compiler. It has been extracted from the bug reports available online. 
Some bugs might have been reported using different mechanisms such as email to developers. There are many more dormant bugs that yet to be found. Therefore, it would not be a representative of all bugs. However, using it in the \KC approach could guide \C{} to trigger several failures.

Second, we only looked at the absolute number of failures. 
We did not check to see if they represent distinct bugs because determining the number of distinct faults is a non-trivial task.

Third, \KC approach divides observations (initial failing \test{}s) into $k$ clusters that are similar within-cluster but dissimilar between-cluster. In data with high dimensions, the value of $k$ is important. The wrong choice of $k$ can push clustering to include dissimilar observations or to exclude similar ones. Unfortunately, there is no well-established method to reach to the right value for $k$. Developers mostly suggest to a fail and trial approach with  multiple  $k$.

Another concern in this paper is that we only did experiments with only two versions of \G{}. As a result, the observed failing \test{} may be fixed or passed by a newer version of \G{}. We run each script for 13 hours and evaluated the \test{}s in this time window. The number of \test{}s and the allocated time may also impact on the evaluation result. Also, the different compilation timeout may affect the number in the result.

\section{Conclusion}
\label{sec:conclution}

Compilers are key software tools for developers to build software. Compiler testing is necessary to ensure the correctness of compiler. In this paper, we have proposed \KC to create a configuration for test generators by processing existing test inputs. 
We experimented this approach on two versions of \G{} compilers and found that the configuration suggested by \KC could trigger several crashes and miscompilation failures in two stable versions of \G{}.

\bibliography{references}
\bibliographystyle{IEEEtranS}

\end{document}